# Finite Element Modelling of Impedance Spectroscopy Data in Composite Electroceramics

Alexander Goncharov[1], Gino Hrkac[1], Julian Dean[1] and Thomas Schrefl[1,2]

[1]*Department of Materials Science and Engineering, University of Sheffield, Mappin Street, Sheffield, S1 3JD, UK*

[2]*Department of Computer Science, St. Poelten University of Applied Science, St. Poelten, Austria*

**Abstract**

A time domain finite element numerical study of impedance spectroscopy in composite electroceramics is presented. The simulations take into account the complexity of the realistic three dimensional granular structure including grains and grain boundaries. Two types of finite elements are used to represent the bulk (tetrahedrons) and the grain boundary (prisms) phases, which allow using extremely high bulk to grain boundary ratios. Simulations at various frequencies reveal a strong influence of the grain shape on the impedance data of the sample, both at low and at high frequencies.

# INTRODUCTION

Originally developed for the solution of problems in solid mechanics, today the Finite Element Method (FEM) is a widely used technique for numerical simulations in many areas of physics and engineering. In this work the Time Domain Finite Element Method (TDFEM) is used for the simulation of impedance spectroscopy data in composite electroceramics. Traditionally, when dealing with harmonic fields, Maxwell equations are transformed into the frequency domain using Fourier transform[1]. In other techniques harmonic fields are represented using complex exponents and the media is described by a complex permittivity [1]. In both methods Maxwell equations are transformed to an identical set of algebraic equations. In those equations the differential operator $\partial/\partial t$ is replaced with the multiplicative factor $i\omega$, which makes them easier to solve numerically. With the general variation of the electric field with time the usage of the time domain method can become advantageous. As an example, a time domain FEM treatment of quasistatic electric problems in the media with frequency dependent permittivity was presented in [2]. The idea of using the FEM for the modeling of impedance spectroscopy data is not new, but the comprehensive treatment of granular three dimensional samples is missing [3]. In this paper, the finite elements method is applied for the simulations of complex electroceramics with three dimensional grains. Simulations have been used to study and explain the influence of the microstructure on the impedance data of electroceramics.

# IMPEDANCE SPECTROSCOPY AND BLM

Impedance spectroscopy (IS) is a widely used methodology to study electrical properties of composite electroceramics [4]. The impedance data can be easily obtained by measuring a time-dependent current response of the sample to the applied oscillating voltage for different frequencies. In general, the impedance is a complex-valued function of the frequency and is given by:

$$Z(\omega) = \frac{V_0 \cos(\omega t + \varphi_V)}{I_0 \cos(\omega t + \varphi_I)} = \text{Re}(Z) + i\text{Im}(z) = |Z|\cos(\theta) + i|Z|\sin(\theta) \quad (1)$$

Where $V_0$ and $I_0$ are amplitudes of the voltage and the current respectively, $\omega$ is the circular frequency, and $\theta = \varphi_V - \varphi_I$ is the phase difference between the driving voltage and the current. For the pure resistor, the phase difference is zero and equation (1) turns to the integral Ohm's law with the real resistance independent of the frequency. For a pure capacitor the phase difference is -90° and for a pure inductor it is equal to 90°. The complex values of these impedances are $Z_C(\omega) = -i/\omega C$ and $Z_L(\omega) = i\omega L$ respectively. For a circuit, which contains all types of elements the phase difference lies between these two numbers and depends on a driving frequency. For example for a resistor and a capacitor connected in series, at low frequencies the total phase difference is close to -90°; there is not enough time to charge the capacitor. But at frequencies for which $\omega \gg 1/RC$ the contribution to the impedance from the capacitor is negligible and the phase shift approaches zero.

In the impedance spectroscopy the impedance is measured over a range of frequencies and the so-called impedance spectrum is obtained. Then the imaginary part is plotted versus the real one on the complex plane forming a Nyquist plot. The frequency, which corresponds to the top of the arc, is given by an inverse of the time constant: $\omega_{top} = 1/RC$. The diameter of the arc is equal to the resistance $R$. Sweeping the frequency allows us to capture features corresponding to different time constants. For example, if the sample has only two phases – grain cores (low $R$) and grain boundaries (high *R*), then the impedance spectrum also has two corresponding arcs. The arc at high frequencies is associated with the grain core properties and the arc at lower frequencies is associated with grain boundaries as in Fig. 1, where frequency is increasing from left to right. Such a close similarity between the impedance spectrums of the RC circuit and electroceramics served as a ground for the brick layer model (BLM). The first and very simple BLM was proposed by Buerle in 1969 and consisted of two parallel RC circuits connected in series [5]. From that other models have been developed and can be found in [4]. The common feature between all these models is that they try to model the AC-response of electroceramics to the harmonic excitation by means of equivalent circuits. The behavior of the time-dependent current in equivalent circuits is described by the integral laws of the quasi-static electromagnetic field in the continuous media. This method proves to be very effective at the microscale (thin grain boundaries), but drastically fails for thick grain boundaries, because real grains do not have a cubic shape [4]. Other techniques such as FEM, should be used instead.

## QUASISTATIC ELECTRIC FIELD

The time evolution of the spatial distribution of the electric field in the sample can be computed by solving the Maxwell's equations in space and time. To simplify our problem we take advantage of the fact that in most electroceramic materials inductive effects can be neglected in favor of capacitive effects. Hence, the quasistatic electric field is curl-free and can be described as a gradient of a scalar electric potential. Since the electric field in the sample changes with time at frequencies comparable to an inverse of the $\tau = 1/RC$, it is necessary to treat conduction and displacement current simultaneously. In the following we will assume that the media is linear, isotropic and there is no time dispersion present, so the electric permittivity is the function of the position only. ($\varepsilon = \varepsilon(\vec{r})$). For such a media the relation between the time-dependent electric displacement and the electric field is a simple relation:

$$\vec{D}(\vec{r},t) = \varepsilon(\vec{r})\vec{E}(\vec{r},t) \qquad (2)$$

The governing equation for the computation of the electric field in the electroceramic sample is obtained from the continuity equation:

$$\frac{\partial \rho}{\partial t} + \nabla \cdot \vec{j} = 0 \qquad (4)$$

which is transformed to:

$$-\nabla \cdot \left( \sigma(\vec{r}) \nabla \varphi(\vec{r},t) + \varepsilon(\vec{r}) \frac{\partial}{\partial t} \nabla \varphi(\vec{r},t) \right) = 0 \qquad (5)$$

In (5) the total current density consists of two parts: the conductive part: $\vec{j}_c = \sigma \vec{E}$ - differential Ohm's law, and the displacement current density: $\vec{j}_d = \partial \vec{D}/\partial t$. The following substitutions have also been made: $\nabla \cdot \vec{E} = \rho/\varepsilon$ and $\vec{E} = -\nabla \varphi$. If the electric field is harmonic and is given at any point in space by $\vec{E}(\vec{r},t) = \text{Re}\left\{ \vec{E}_0(\vec{r}) e^{i\varphi(\vec{r})} e^{i\omega t} \right\}$, then for the total current density in the unit volume is expressed as follows

$$\vec{J}(\vec{r},t) = Z^{-1}(\vec{r},\omega)\vec{E}(\vec{r},t)$$
$$Z^{-1}(r,\omega) = \sigma(\vec{r}) + i\omega\varepsilon(\vec{r}) = \frac{1}{Z_R(\vec{r})} + \frac{1}{Z_C(\vec{r},\omega)} \qquad (6)$$

Two parts of $Z^{-1}(\vec{r},\omega)$ can be associated with the real ($Y_R$ - conductance) and the imaginary ($Y_C$ -susceptance) parts of the admittance. The resistance and the capacitance of the unit volume are equal to $\sigma^{-1}$ and $\varepsilon$ respectively. Equation (6) differs from (1), because it relates the local electric field and the local current density, which is caused by this local field. If this relation is known at any point inside the sample, one can easily compute $\vec{J}(\vec{r},t)$ and then $Z(\vec{r},\omega)$ by integrating over the whole sample. Equation (5) for the electric potential is solved for every frequency using FEM with the following boundary conditions:

(1) Fixed potential values at electrodes-air boundaries. (Dirichlet)

(2) Free sides satisfy: $\left(\sigma\nabla\varphi + \varepsilon\frac{\partial}{\partial t}\nabla\varphi\right)\cdot\vec{n} = 0$. (Neumann).

The unknown solution for the potential $\varphi(\vec{r},t)$ is approximated using finite element shape functions: $\varphi(\vec{r},t) = \sum \varphi_j(t)\psi_j(\vec{r})$, where $N$ is a number of nodes in the mesh, and $\psi_j(\vec{r})$ is a shape function associated with the $j^{th}$ node. By applying the Galerkin scheme with the implicit time discretization, equation (5) is transformed to the following matrix equation:

$$\left(K_\sigma + \frac{1}{\Delta t}K_\varepsilon\right)\varphi^{n+1} = \frac{1}{\Delta t}K_\varepsilon\varphi^n \qquad (7)$$

where $K_\sigma$ and $K_\varepsilon$ are the conductivity and permittivity stiffness matrices, and $\varphi^n$ is the vector with nodal values of the potential at the $n^{th}$ time step. Using computed values for the electric potential at every time step, the local electric field and current can be found at every time step. The current flowing through the electrode for a certain frequency is computed as a following integral over the electrode surface $S_{El}$:

$$I(t) = \iint_{S_{El}} \vec{J}(\vec{r},t)\cdot\vec{n}\,dA \qquad (8)$$

## RESULTS AND DISCUSSION

To verify our approach we simulated a model representing a serial RC circuit. This model was chosen because it can be exactly fitted using an equivalent circuit theory. A 100 nm long cylinder with a base radius of 25 nm was used, which consisted of two parts: one representing a pure resistance ($\sigma = 1\times10^{-4}$, $\varepsilon = 0$) and the other a pure capacitance ($\sigma = 0$, $\varepsilon = 1\times10^{-9}$). For these dimensions the values for $R$ and $C$ are $2.546\times10^{11}$ Ohm and $3.93\times10^{-17}$ F respectively. Simulations of the impedance spectrum were performed for the frequencies ranging from 10 Hz to 120 MHz. Computed values of $|Z|$ (circles) and the fitting (solid line) are shown in Fig. 3. Using equivalent $RC$ circuit fitting, the values for $R$ and $C$ ware found to be $2.55\times10^{11}$ and $3.91\times10^{-17}$ respectively. These values are in a good agreement with those used in our simulations.

BLM has been reasonably successful for the interpretation of the impedance spectroscopy, but real ceramics are significantly different from that simplified model. Not only the real samples have a complex microstructure, but also their local properties can change from grain to grain. FEM offers a great flexibility in performing simulations of such complex systems. In our simulations, realistic 3D granular structures are modeled using a 3D Voronoi tessellation. Example of a 3D ceramics model with irregular grains is shown in Fig. 4. Grains are discretized using tetrahedral finite elements and are assigned their local values for conductivity and permittivity. These parameters can be easily changed to amend the behavior of the whole sample. For example, the local parameters can change from grain to grain, can have gradual transitions, or be temperature dependent. The space between grains is discretized using prismatic elements (not shown). The usage of prismatic elements between the grains allows us to use extremely thin grain boundaries (aspect ratios in the order of 100 and more). These thin grain boundaries are extremely difficult to discretize using only tetrahedral elements and in most cases the preprocessing software will fail to mesh the region. Both tetrahedral and prismatic elements adopt a linear approximation to the unknown potential inside of each element. Details of the usage of the prismatic elements in conjunction with tetrahedral elements can be found in [6].

In the following we will present two case studies of how the microstructure influences the impedance spectroscopy data: effect of the shape and the number of grains. To study the effect of the grain shape on the impedance spectroscopy data, two samples were chosen. Both have the same cubic shape with an edge length equal to 1 micrometer. One model has 125 cubic grains and the other has 125 irregularly shaped grains. The aspect ratio between the grains size and the grain boundaries was set to 10. Electrical conductivities for the grain and grain boundary phases were $1\times10^{-4}$ S/m and $1\times10^{-6}$ S/m respectively. The value of the permittivity of $1\times10^{-10}$ F/m was used for the whole sample. Simulations for the frequency range from 10 to $3\times10^6$ Hz revealed that the microstructure can significantly influence the impedance data. Fig. 5 shows impedance spectrums for two samples (solid line corresponds to cubic grains, dashed – to irregular grains. As can be seen from the results in Fig. 5, the shape of the grains and grain boundaries play an important role both at low and at high frequencies. However this effect is more pronounced at lower frequencies, where it is attributed to the grain boundary phase. The model with cubic grains tends to underestimate the impedance of the sample with the same dimensions and number of grains. It can be concluded from the simulations that in addition to the aspect ratio between the bulk and grain boundary phases, the shape of the grains is also important for the modeling of impedance spectroscopy data. In the second case, the influence of the number of grains on the impedance of the sample has been investigated. In numerical simulations involving real data it is important to know how many grains are needed to simulate and predict experimental results. For this purpose, the impedance data for three 1 micrometer cubic models with irregular grains was computed. The number of grains used in the three samples are 125, 512 and 729. The higher number of grains guarantees that there are more grain boundaries and there is a more complicated microstructure inside the sample. Approximated grain sizes varied from 200 nm to 110 nm for these numbers of grains. Fig. 6 shows impedance spectrums for those three samples, where it is clearly seen that there is no difference in the response at high frequencies. Hence, in contrast to the results of the previous case shown in Fig. 5, the number of grains used in simulations does not influence the response of the bulk phase as soon as they all have irregular shapes. However, at low frequencies, the number of grains changes the impedance of the sample. Analysis of the distribution of the electric potential and the electric field inside the samples helps to understand this sort of behavior. In Fig. 7 there are two contour plots of the electric potential in a cross-section through the sample with 125 grains at low (A) and at high (B) frequency. Fig. 8 shows local electric fields for those electric potentials for the same frequency values. The results show that at low frequencies (Fig. 8A), the number of grains as well as their shape is extremely important for the local current flows. At high frequencies the grain boundaries do not have enough time to build up a charge, and grain boundaries do not affect the electric field, which is now pointed along Z direction (Fig. 8B). By increasing the number of grains, the number of grain boundaries, their shapes and orientations changes as well. This leads to the different simulated impedance data. Since we compute the average response of the whole sample, the number of grains in the sample does not change the results of the simulations indefinitely – there is no significant change in the computed impedance data for the samples having 512 and 729 grains. This result suggests that the model, which has approximately 500 grains in total, can be used for the simulation of the experimental impedance data.

## CONCLUSION

A time domain finite element software was used to investigate the influence of the microstructure on the impedance data modeling. The simulation results show that the shape of the grains can significantly alter the impedance spectroscopy data at low frequencies. The simulated data also shows that the aspect ratio is not the only parameter, which can significantly impact the behaviour of the sample. Numerical simulations performed on cubic and non-cubic grains with the same aspect ratio clearly demonstrate this impact.

## References

[1] J. D. Jackson, Classical Electrodynamics, 3d ed., John Wiley & Sons, Inc.

[2] K. Preis, O. Bíró, P. Supancic, I. Tičar, G. Matzenauer, *IEEE Trans. Mag.*, **40** (2), 1302-1305, (2004)

[3] J. Fleig and J. Maier, *J. Am. Cheram. Soc.*, **82** [12], 3485-3493, (1999)

[4] N.J. Kindner, N.H. Perry and T.O. Mason, *J. Am. Ceram. Soc.*, **91** [6] 1733-1746, (2008)

[5] J. E. Bauerle, *J. Phys. Chem, Solids*, **30**, 2657-2670, (1969)

[6] Ch. Guérin, G. Tanneau, *IEEE Trans. Mag.*, **30** (5), 2885-2888, (1994)

**Figures captions**

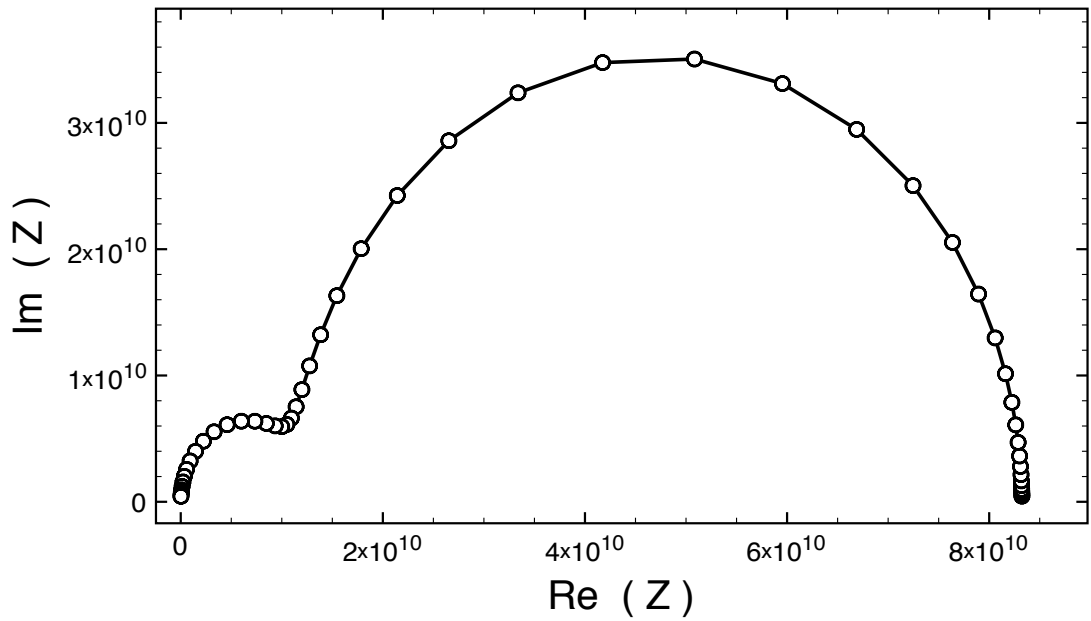

Fig. 1. A dual arc impedance spectrum of a double phase electroceramic sample. Each arc can be modeled as a parallel resistor-capacitor circuit.

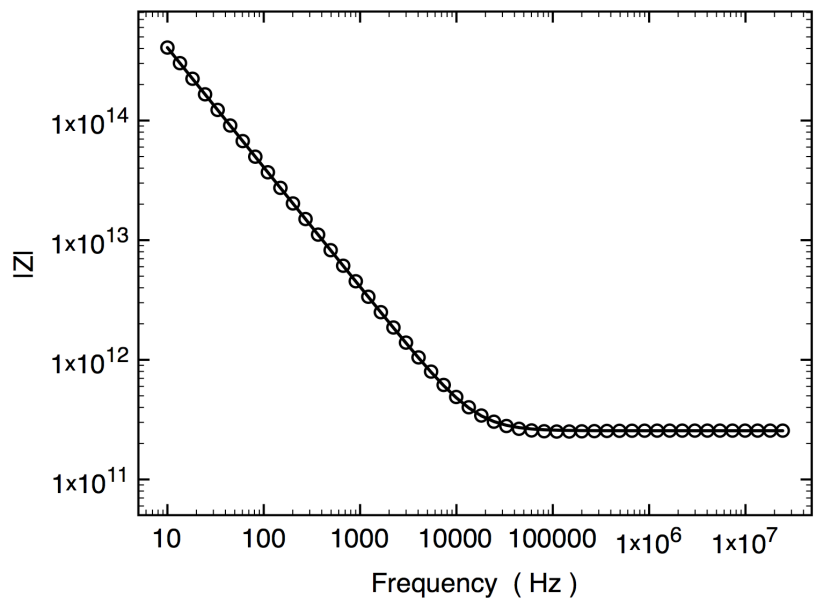

Fig. 2. Frequency dependence of the absolute value of the impedance. Solid line Represents the equivalent circuit fitting with $R=2.55\times10^{11}$ and $C=3.91\times10^{-17}$

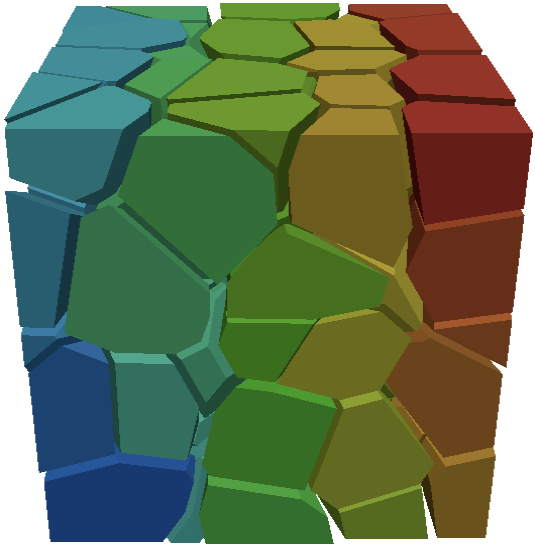

Fig. 3. (color online)The model of the electroceramic sample with irregular granular structure. Grains are discretized using tetrahedral finite elements. Interstitial spaces between grains are grain boundaries and are filled with prismatic elements.

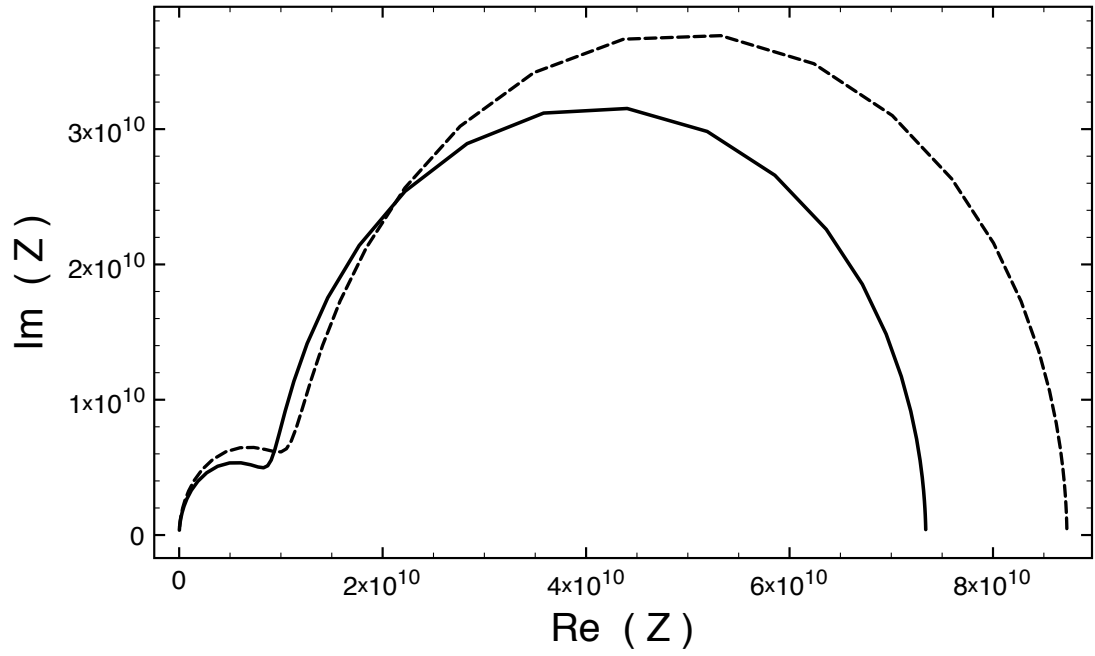

Fig. 4. An impedance spectrum for cubic grains (solid line) and irregular grains (dashed line).

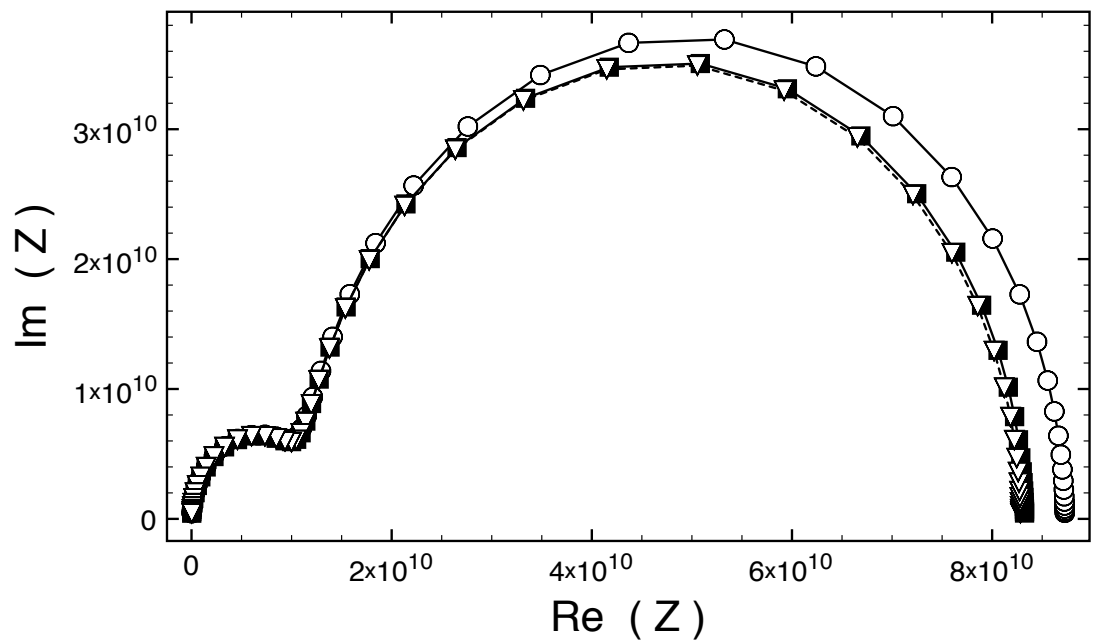

Fig. 5. An impedance spectrum for the different number of irregular grains. Circles – 125, black squares - 512, triangles – 729.

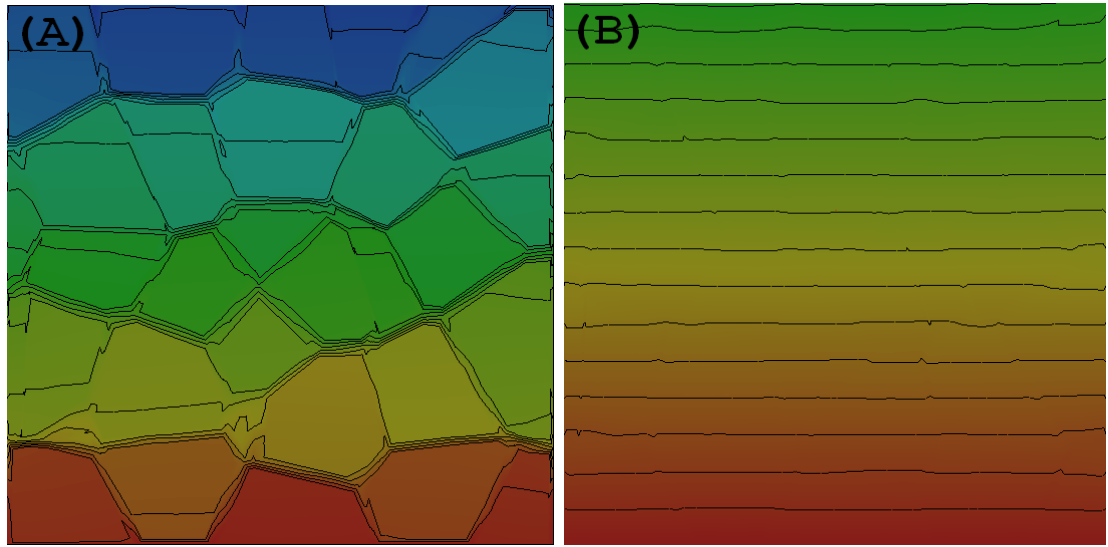

Fig. 6. (color online) Computed electric potential at low frequencies (A) and at high frequencies (B). The effect of grain boundaries disappears at high frequencies. A cross-section through the middle of the sample is shown.

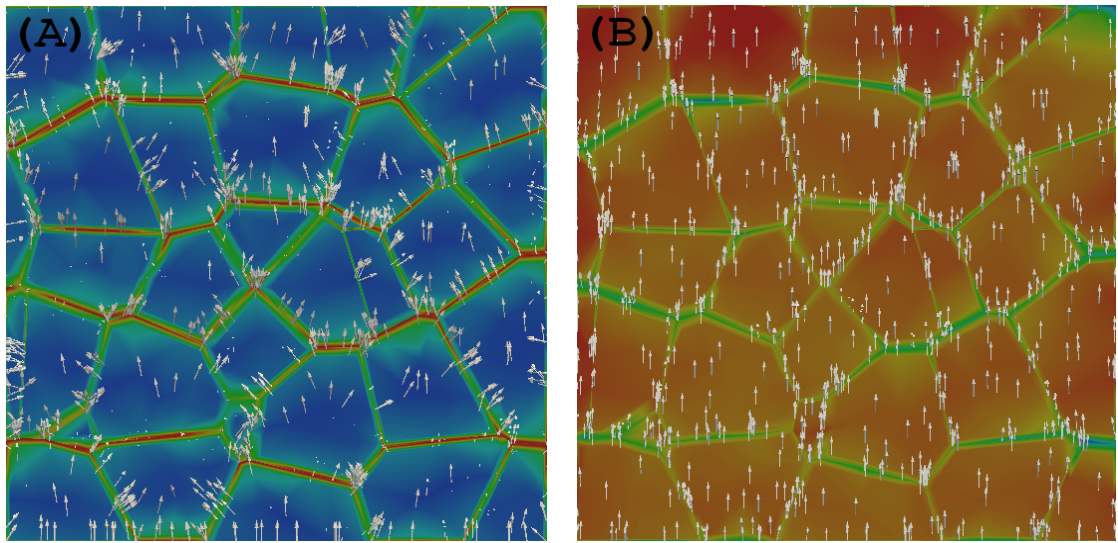

Fig. 7. (color online) Computed local electric field for the potential shown in Fig. 6. Low frequencies data shown in (A) and high frequencies data is in (B). A cross-section through the middle of the sample is shown.